\documentclass[12pt, a4paper]{article}
\usepackage{epsf}
\usepackage{cite}
\usepackage{amsmath,amssymb}
\input{colordvi.tex}
\usepackage[usenames,dvipsnames]{color}
\usepackage{graphicx}
\usepackage{ascmac}
\usepackage{hyperref}
\usepackage{tikz}
\usepackage{tikz-feynhand}
\usepackage[subrefformat=parens]{subcaption}

\begin{document}

\begin{titlepage}

\begin{center}
\hfill TU-1223\\
\hfill KEK-QUP-2024-0005
\vskip 1.0in

{\Large \bf
Effects of gravitational particle production on  \\ \vspace{2.5mm} 
Higgs portal dark matter
}

\vskip .5in

{\large
Soichiro Izumine$^{(a)}$ and
Kazunori Nakayama$^{(a,b)}$
}

\vskip 0.5in

$^{(a)}${\em 
Department of Physics, Tohoku University, Sendai 980-8578, Japan
}

\vskip 0.2in

$^{(b)}${\em 
International Center for Quantum-field Measurement Systems for Studies of the Universe and Particles (QUP), KEK, 1-1 Oho, Tsukuba, Ibaraki 305-0801, Japan
}

\end{center}
\vskip .5in

\begin{abstract}
The gravitational interaction is ubiquitous and the effect of gravitational particle production necessarily contributes to the dark matter abundance. A simple candidate of dark matter is a scalar particle, whose only renormalizable interaction is the Higgs portal coupling. We show that the abundance of Higgs portal dark matter is significantly affected by the gravitational production effect. In particular, the gravitational production from the coherently oscillating inflaton field during the reheating often gives dominant contribution.
\end{abstract}

\end{titlepage}

\tableofcontents

\section{Introduction}

Gravitational production of particles in an expanding universe~\cite{Parker:1969au} may have a lot of impacts on cosmology.
In inflationary universe~\cite{Starobinsky:1980te,Guth:1980zm,Sato:1981qmu,Linde:1981mu,Albrecht:1982wi}, the gravitational particle production at the transition from the de Sitter universe to the matter- or radiation-dominated universe can produce particles as heavy as the Hubble scale of inflation $H_{\rm inf}$~\cite{Ford:1986sy} and it may be a source of dark matter (DM) of the present universe~\cite{Chung:1998zb,Kuzmin:1998kk,Chung:2011ck}. 
The gravitational production process has a special meaning compared with other production mechanisms in the sense that it is ubiquitous. Unless conformally coupled to the gravity, any particle is subject to the gravitational production.
Recently the gravitational production effects draw lots of attention after it has been pointed out that the inflaton coherent oscillation contributes to the gravitational production during the reheating era~\cite{Ema:2015dka,Ema:2016hlw,Ema:2018ucl,Ema:2019yrd,Chung:2018ayg}.
As opposed to the conventional wisdom, it became clear that the gravitational production can produce particles with mass heavier than the Hubble scale, as far as it is lighter than the inflaton mass.
Intuitively this process can be interpreted as the perturbative inflaton annihilation through the gravitational interaction as already pointed out in Refs.~\cite{Ema:2015dka,Ema:2016hlw,Tang:2017hvq,Ema:2018ucl,Ema:2019yrd,Chung:2018ayg}. Several phenomenological implications have been discussed in Refs.~\cite{Ahmed:2020fhc,Mambrini:2021zpp,Clery:2021bwz,Clery:2022wib,Haque:2021mab,Haque:2022kez,Kaneta:2022gug,Ahmed:2022tfm}.
On the other hand, the Standard Model particles in thermal bath also gravitationally produces the DM particles through the s-channel graviton exchange and it can also account for the DM abundance~\cite{Garny:2015sjg,Tang:2016vch,Garny:2017kha,Tang:2017hvq}.

In these studies, it has been assumed that the DM only has a gravitational interaction. While it is a very simple assumption, the four-point coupling between the DM and the Higgs field (the Higgs portal coupling) of the from $\mathcal L\sim -\lambda \chi^2 |\mathcal H|^2/2$ is not prohibited by the symmetry argument if the DM $\chi$ is a scalar field, where $\mathcal H$ is the Standard Model Higgs field and $\lambda$ is the coupling constant.
For sizable coupling $\lambda$, the DM is thermalized through the Higgs portal coupling and the DM abundance is determined by the freezeout scenario~\cite{Silveira:1985rk,McDonald:1993ex} (see Ref.~\cite{Arcadi:2019lka} for a review). For such a case the gravitational effect is negligible and we do not consider this scenario.
For very small coupling $\lambda$, on the other hand, DM is never thermalized but still produced gradually through the Higgs portal coupling and it can account for the observed DM abundance~\cite{Kolb:2017jvz,Chianese:2020yjo,Aoki:2022dzd}.
In this paper we fully include the effect of Higgs portal coupling and the gravitational coupling for the scalar DM production mechanisms and evaluate the DM abundance. 

Let us comment on past related works. In the most works treating the gravitational production of DM, the Higgs portal coupling has been neglected by  assuming that the DM has only the gravitational interaction. There are a few works that considered both couplings. 
Ref.~\cite{Kolb:2017jvz} considered the production of superheavy DM through the Higgs portal coupling and the gravitational production during inflation,\footnote{
	Creation of long wave quantum fluctuations of light field during inflation is regarded as a gravitational production. Ref.~\cite{Kolb:2017jvz} actually considered such an effect, assuming that it results in the misalignment production with an initial scalar amplitude of $\sim H_{\rm inf}$. We do not consider misalignment contribution, since it is highly model dependent. See Sec.~\ref{sec:other}. 
} while the gravitational production from the Standard Model plasma and inflaton coherent oscillation are not taken into account. 
Ref.~\cite{Chianese:2020yjo} considered the effect of Higgs portal coupling and gravitational production from the Standard Model plasma, but neglected that from the inflaton coherent oscillation.
Ref.~\cite{Aoki:2022dzd} included all these effects in a particular model of the Higgs-$R^2$ inflation~\cite{Salvio:2015kka,Salvio:2016vxi,Ema:2017rqn,Gorbunov:2018llf,Gundhi:2018wyz} but they mostly focused on the case of vanishing Higgs portal coupling or the conformal coupling where the effect of inflaton oscillation is negligible.

This paper is organized as follows. In Sec.~\ref{sec:prod}, the DM production rates from various processes are summarized, including the production from the Higgs portal coupling and gravitational production from the Standard Model particles and the coherently oscillating inflaton.
In Sec.~\ref{sec:dm} the resulting DM abundance is evaluated in both semi-analytic and numerical way and present explicitly parameters to reproduce the observed DM abundance.
In Sec.~\ref{sec:other} we consider possible other effects that may affect the DM abundance.
We conclude in Sec.~\ref{sec:con}.
In App.~\ref{app:th} we give useful formulae for calculating thermally averaged cross section for various processes.
In App.~\ref{app:inf} several inflation models consistent with the current observation are listed for concreteness, although we do not necessarily require details of the inflation model for calculating the DM abundance.

\section{Dark matter production rate} \label{sec:prod}

\begin{figure}[tbp]
  \begin{center}
    \begin{tabular}{ccc}
      \begin{minipage}[b]{0.3\hsize}
        \centering
        \begin{tikzpicture}
          \begin{feynhand}
            \vertex (o) at (0,0);
            \vertex (h1) [particle, above left = of o] {$\mathcal H$};
            \vertex (h2) [particle, below left = of o] {$\mathcal H$};
            \vertex (chi1) [particle, above right = of o] {$\chi$};
            \vertex (chi2) [particle, below right = of o] {$\chi$};
            \propag [plain, mom=$p_1$] (h1) to (o);
            \propag [plain, mom=$p_2$] (h2) to (o);
            \propag [sca, mom'=$p_3$] (o) to (chi1);
            \propag [sca, mom'=$p_4$] (o) to (chi2);
          \end{feynhand}
        \end{tikzpicture}
        \subcaption{}
        \label{fig:higgs-portal}
      \end{minipage}
      &
      \begin{minipage}[b]{0.3\hsize}
        \centering
        \begin{tikzpicture}
          \begin{feynhand}
            \vertex (o) at (0,0);
            \vertex (phi1) [particle, above left = of o] {SM};
            \vertex (phi2) [particle, below left = of o] {SM};
            \vertex (oo) [right = 2cm of o];
            \vertex (chi1) [particle, above right = of oo] {$\chi$};
            \vertex (chi2) [particle, below right = of oo] {$\chi$};
            \propag [plain, mom=$p_1$] (phi1) to (o);
            \propag [plain, mom'=$p_2$] (phi2) to (o);
            \propag [bos, mom=$k$] (o) to (oo);
            \propag [sca, mom=$p_3$] (oo) to (chi1);
            \propag [sca, mom'=$p_4$] (oo) to (chi2);
          \end{feynhand}
        \end{tikzpicture}
        \subcaption{}
        \label{fig:SM-to-chi}
      \end{minipage}
      &
      \begin{minipage}[b]{0.3\hsize}
        \begin{tikzpicture}
          \begin{feynhand}
            \vertex (o) at (0,0);
            \vertex (phi1) [particle, above left = of o] {$\phi$};
            \vertex (phi2) [particle, below left = of o] {$\phi$};
            \vertex (oo) [right = 2cm of o];
            \vertex (chi1) [particle, above right = of oo] {$\chi$};
            \vertex (chi2) [particle, below right = of oo] {$\chi$};
            \propag [sca, mom=$p_1$] (phi1) to (o);
            \propag [sca, mom'=$p_2$] (phi2) to (o);
            \propag [bos, mom=$k$] (o) to (oo);
            \propag [sca, mom=$p_3$] (oo) to (chi1);
            \propag [sca, mom'=$p_4$] (oo) to (chi2);
          \end{feynhand}
        \end{tikzpicture}
        \subcaption{}
        \label{fig:phi-to-chi}
      \end{minipage}
    \end{tabular}
  \end{center}
  \caption{Feynman diagrams for DM production processes. The wavy line represents the graviton.}
\end{figure}
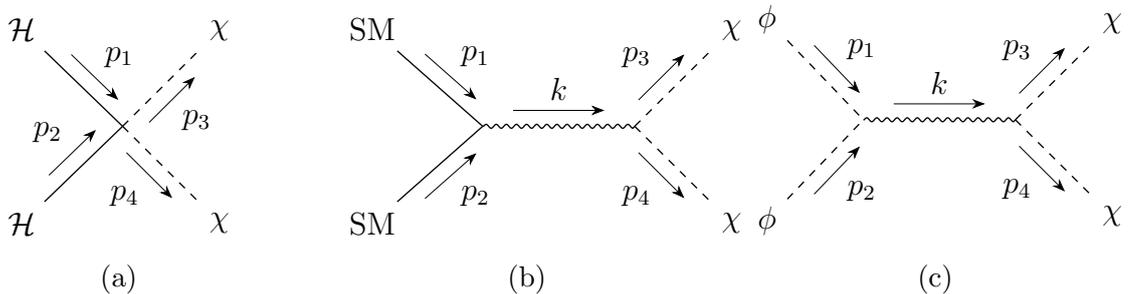

The total action we consider is
\begin{align}
	S = \int d^4x \sqrt{-g}\left(-\frac{M^2_{\rm Pl}}{2} R + \mathcal L_{\phi} +  \mathcal L_{\chi} + \mathcal L_{\rm HP} +  \mathcal L_{\rm SM} \right), 
	\label{action}
\end{align}
where $M_{\rm Pl}$ is the reduced Planck scale, $R$ the Ricci curvature, $\mathcal L_\phi$ is the Lagrangian for the inflaton, $\mathcal L_\chi$ is the Lagrangian for the DM, $\mathcal L_{\rm SM}$ is the Standard Model Lagrangian and $\mathcal L_{\rm HP}$ represents the Higgs portal interaction of the DM: 
\begin{align}
	&\mathcal L_\phi = \frac{1}{2}g^{\mu\nu} \partial_\mu\phi \partial_\nu\phi - V(\phi),\\
	&\mathcal L_\chi = \frac{1}{2}g^{\mu\nu} \partial_\mu\chi \partial_\nu\chi - \frac{1}{2}m_\chi^2\chi^2,\\
	&\mathcal L_{\rm HP} = -\frac{\lambda}{2} |\mathcal H|^2 \chi^2,
\end{align}
where $\phi$ is the inflaton, $\chi$ is a real scalar DM candidate, $\mathcal H$ is the Standard Model Higgs doublet.
We assign the $Z_2$ symmetry under which only $\chi$ flips its sign, which ensures the absolute stability of $\chi$ so that it can take a role of DM.
From this action, we can derive the graviton interaction by expanding the metric around the Minkowski metric $\eta_{\mu\nu}$ as $g_{\mu\nu} = \eta_{\mu\nu} + (2/M_{\rm Pl})h_{\mu\nu}$~\cite{Maggiore:2007ulw}
\begin{align}
	\mathcal L = \frac{1}{M_{\rm Pl}} h_{\mu\nu} T^{\mu\nu},
\end{align}
where $T^{\mu\nu}$ is the energy-momentum tensor of matter and $h_{\mu\nu}$ denotes the canonically normalized graviton.
The Feynman rules for the vertex and the graviton propagator is found e.g. in Ref.~\cite{Clery:2021bwz}. The graviton propagator in the Feynman gauge is~\cite{Maggiore:2007ulw}
\begin{align}
	\Pi^{\mu\nu\rho\sigma} = \frac{i}{2k^2}(\eta^{\mu\rho}\eta^{\nu\sigma}+\eta^{\mu\sigma}\eta^{\nu\rho}-\eta^{\mu\nu}\eta^{\rho\sigma}),
\end{align}
and the vertex for a massless scalar, chiral fermion and vector boson with the graviton is
\begin{align}
	&V^{\rm (scalar)}_{\mu\nu}= \frac{-i}{2M_{\rm Pl}}(p_{1\mu}p_{2\nu}+p_{2\mu}p_{1\nu}-\eta_{\mu\nu}p_1\cdot p_2),\\
	&V^{\rm (fermion)}_{\mu\nu}= \frac{-i}{4M_{\rm Pl}}\overline{v_L}(p_2) \left[ \gamma_\mu(p_1-p_2)_\nu + \gamma_\nu(p_1-p_2)_\mu \right] u_L(p_1),\\
	&V^{\rm (vector)}_{\mu\nu}= \frac{-i}{2M_{\rm Pl}}\left[ 
		\epsilon_2^*\cdot\epsilon_1(p_{1\mu}p_{2\nu}+p_{2\mu}p_{1\nu}) 
		+ p_1\cdot p_2 (\epsilon_{1\mu}\epsilon^*_{2\nu} +\epsilon_{1\nu}\epsilon^*_{2\mu}  )
		-\epsilon_2^*\cdot p_1(p_{2\mu}\epsilon_{1\nu}+p_{2\nu}\epsilon_{1\mu})   \right. \nonumber\\
		&\left.~~~~~~~~~~~~  -\epsilon_1\cdot p_2(p_{1\mu}\epsilon^*_{2\nu}+p_{1\nu}\epsilon^*_{2\mu}) 
		+\eta_{\mu\nu}( (p_1\cdot\epsilon^*_2)(p_2\cdot\epsilon_1)-(p_1\cdot p_2)(\epsilon_1\cdot\epsilon_2^*) )
 	\right],
\end{align}
where $\gamma^\mu$ is the Dirac gamma matrices, $u_L$ and $v_L$ are the left-handed spinors for an incoming particle and anti-particle and $\epsilon_1, \epsilon_2$ are the polarization tensors for vector boson, following the notation of Ref.~\cite{Peskin:1995ev}.
Below we summarize the thermally averaged cross section or the production rate for DM production processes. 
Corresponding diagrams are shown in Fig.~\ref{fig:higgs-portal}, \ref{fig:SM-to-chi}, \ref{fig:phi-to-chi} for the production through Higgs portal coupling, gravitational production from the Standard Model particles and inflaton, respectively.
Some technical details to evaluate the cross section are summarized in App.~\ref{app:th}.

\subsection{Production from Higgs portal coupling}

First let us consider the production from the Higgs portal coupling (Fig.~\ref{fig:higgs-portal}). We assume that the Higgs particles are thermalized with the temperature $T$. 
By noting that the Higgs field $\mathcal H$ contains four real degrees of freedom at the symmetric phase and using the formula presented in App.~\ref{app:th}, thermally averaged cross section for the Higgs portal coupling is given by
\begin{align}
    \left<\sigma v\right>^{\rm (HP)}_{\chi\chi\to \mathcal H \mathcal H} = \frac{\lambda^2}{16\pi m_\chi^2}\left(\frac{K_1(m_\chi/T)}{K_2(m_\chi/T)}\right)^2
    \simeq \begin{cases}
        \displaystyle \frac{\lambda^2}{16\pi m_\chi^2} ~~ {\rm for}~m_\chi \gg T\\
        \displaystyle \frac{\lambda^2}{64\pi T^2} ~~ {\rm for}~m_\chi \ll T
    \end{cases},
    \label{sigmaHP}
\end{align}
where $K_n$ denotes the modified Bessel function of the second kind with order $n$.
Note that we have actually evaluated the DM annihilation cross section $\left<\sigma v\right>^{\rm (HP)}_{\chi\chi\to \mathcal H \mathcal H}$, not the production cross section itself $\left<\sigma v\right>^{\rm (HP)}_{\mathcal H \mathcal H \to\chi\chi}$. They are related through the detailed balance condition, as shown in the next section.
This is why the cross section (\ref{sigmaHP}) does not receive a Boltzmann suppression factor $e^{-m_\chi/T}$ for $m_\chi \gg T$. Instead, the Boltzmann suppression factor appears in the hypothetical thermal abundance of $\chi$, $n_\chi^{\rm (th)}$ in Eq.~(\ref{dotnchi}).
The same caution applies to the next case.

\subsection{Production from Standard Model particles through gravity}

Next let us consider the production from the Standard Model particles in thermal bath through gravity (Fig.~\ref{fig:SM-to-chi}). 
The Standard Model particles include spin 0 particles (Higgs boson), spin $\frac{1}{2}$ particles (quarks and leptons) and spin 1 particles (gauge bosons).
Below we neglect the mass of Standard Model particles since we are considering the symmetric phase at high temperature, where the electroweak symmetry is restored. Technical details are found in App.~\ref{app:th}.
The cross section for spin 0 particle, or a real scalar, is given by
\begin{align}
    \left<\sigma v\right>^{\rm (grav)}_{\chi\chi\to 00} &= \frac{m_\chi^2}{320\pi M_{\rm Pl}^4}
    \left(\frac{3K_1^2}{K_2^2} + \frac{4T}{m_\chi}\frac{K_1}{K_2} + \frac{8T^2}{m_\chi^2} + 2\right)  \label{sigmagrav_00} \\
    &\simeq \begin{cases}
        \displaystyle \frac{m_\chi^2}{64\pi M_{\rm Pl}^4} ~~ {\rm for}~m_\chi \gg T\\
        \displaystyle \frac{T^2}{40\pi M_{\rm Pl}^4} ~~ {\rm for}~m_\chi \ll T
    \end{cases}.
\end{align}
Here and in what follows the argument of the Bessel functions, which is $m_\chi/T$, is omitted for simplicity.
The cross section for spin $\frac{1}{2}$ particle, or a chiral fermion, and spin 1 particle, or a massless vector boson, are given by
\begin{align}
    4\left<\sigma v\right>^{\rm (grav)}_{\chi\chi\to \frac{1}{2}\frac{1}{2}} =
    \left<\sigma v\right>^{\rm (grav)}_{\chi\chi\to 11} 
    &=\frac{m_\chi^2}{120\pi M_{\rm Pl}^4}
    \left(\frac{K_1^2}{K_2^2} + \frac{3T}{m_\chi}\frac{K_1}{K_2} + \frac{6T^2}{m_\chi^2} -1 \right)  \label{sigmagrav_11} \\
    &\simeq \begin{cases}
        \displaystyle \frac{T^2}{16\pi M_{\rm Pl}^4} ~~ {\rm for}~m_\chi \gg T\\
        \displaystyle \frac{T^2}{20\pi M_{\rm Pl}^4} ~~ {\rm for}~m_\chi \ll T
    \end{cases}.
\end{align}
It is seen that, in the low temperature limit $T\ll m_\chi$, the scalar production rate remains finite while the fermion and vector boson production rate vanishes as far as the mass of fermion and vector boson is neglected.
It is consistent with the results in Refs.~\cite{Ema:2015dka,Ema:2016hlw,Ema:2018ucl,Ema:2019yrd} where the production from non-relativistic inflaton oscillation has been considered. If the momentum of the initial state scalar is negligibly small, the only gravitational effect of the initial state is to affect the conformal factor and hence non-conformal spin 0 particles are produced while there are no fermion and vector boson production since they are conformal in the massless limit.

Taking account of the Standard Model degrees of freedom, the total cross section is given by
\begin{align}
    \left<\sigma v\right>_{\chi\chi\to {\rm SM}\,{\rm SM}} = \left<\sigma v\right>^{\rm (HP)}_{\chi\chi\to \mathcal H \mathcal H} + 4\left<\sigma v\right>^{\rm (grav)}_{\chi\chi\to 00}+45\left<\sigma v\right>^{\rm (grav)}_{\chi\chi\to \frac{1}{2}\frac{1}{2}}+12\left<\sigma v\right>^{\rm (grav)}_{\chi\chi\to 11}.
\end{align}

\subsection{Production from inflaton through gravity}

Finally we consider the production from the inflaton coherent oscillation.
The inflaton coherent oscillation gravitationally induces the oscillation of the cosmic scale factor $a(t)$ as~\cite{Ema:2015dka}
\begin{align}
	a(t) \simeq \left<a(t)\right> \left(1-\frac{\phi^2-\left<\phi^2\right>}{M_{\rm Pl}^2}\right),
\end{align}
which in turn leads to the production of $\chi$ particles.
The particle production rate is calculated as~\cite{Ema:2015dka,Ema:2016hlw,Ema:2018ucl,Chung:2018ayg}
\begin{align}
	R_{\phi\phi\to\chi\chi} = \frac{27}{256\pi} H^4,
	\label{Rgrav}
\end{align}
for $T \gtrsim T_{\rm R}$, where $H$ is the Hubble parameter. For lower temperature $T \ll T_{\rm R}$, the inflaton coherent oscillation disappears exponentially. This may be interpreted as the perturbative inflaton annihilation to DM as in Fig.~\ref{fig:phi-to-chi}.
As will be explained in the next section, this production is most dominant at earlier epoch (i.e. just after the end of inflation), so we simply neglect it for $T < T_{\rm R}$.

Note that the ``conventional'' gravitational production along the line of Refs.~\cite{Ford:1986sy,Chung:1998zb} is the same order at the transition epoch from de Sitter to the matter- or radiation-dominated universe if $m_\chi \lesssim H_{\rm inf}$.
However, it is not effective for $m_\chi \gg H_{\rm inf}$. On the other hand, the gravitational production from the oscillating inflaton is effective even for $H_{\rm inf} < m_\chi < m_\phi$.
Since $H_{\rm inf} \ll m_\phi$ is natural in many inflation models (see App.~\ref{app:inf}), taking account of the effect from the oscillating inflaton is phenomenologically very important. 
In addition, the ``conventional'' gravitational production in the case of $m_\chi < H_{\rm inf}$ is often hindered by the misalignment contribution, since in this case the long wave quantum fluctuations develop during inflation, which results in homogeneous condensate of the scalar field $\chi$. See Sec.~\ref{sec:mis}.


\section{Dark matter abundance} \label{sec:dm}

\subsection{Thermal history and Boltzmann equation}

We assume the standard thermal history that the inflaton behaves as non-relativistic matter in the reheating phase until it decays into radiation. The evolution of the inflaton energy density $\rho_\phi$ and radiation energy density $\rho_r$ is described by the following equation:
\begin{align}
    &\dot\rho_\phi + 3H \rho_\phi = - \Gamma_\phi \rho_\phi,\\
    &\dot\rho_r + 4H \rho_r = \Gamma_\phi \rho_\phi,
\end{align}
where $\Gamma_\phi$ denotes the inflaton decay rate. It is related to the reheating temperature $T_{\rm R}$ as $T_{\rm R}=\left(\frac{90}{\pi^2 g_*}\right)^{1/4}\sqrt{\Gamma_\phi M_{\rm Pl}}$.
The temperature before the completion of the reheating is given by $T \sim (T_{\rm R}^2 H M_{\rm Pl})^{1/4}$. 
Thus the maximum temperature after inflation is roughly given by $T_{\rm max} \sim (T_{\rm R}^2 H_{\rm inf} M_{\rm Pl})^{1/4}$, where $H_{\rm inf}$ is the Hubble scale at the end of inflation.

The Boltzmann equation for the evolution of the DM number density $n_\chi$ is given by
\begin{align}
	\dot n_\chi + 3H n_\chi = -\left<\sigma v\right>_{\chi\chi\to {\rm SM}\,{\rm SM}}\left(n_\chi^2 - n_\chi^{\rm (th) 2}\right) + R_{\phi\phi\to\chi\chi}.
	\label{dotnchi}
\end{align}
The superscript (th) means that the number density is evaluated as if it is thermalized. 
The first term in the right hand side represents the DM annihilation effect. It would be important in the freeze-out scenario, but it is negligible in our scenario. The second term represents the DM production from the Standard Model plasma. Note that we have used the detailed balance condition: $\left<\sigma v\right>_{{\rm SM}\,{\rm SM}\to\chi\chi} n_{\rm SM}^{\rm (th) 2} =\left<\sigma v\right>_{\chi\chi\to {\rm SM}\,{\rm SM}} n_\chi^{\rm (th) 2}$.

\subsection{Dark matter abundance: analytic estimate}

Here we roughly estimate the abundance of $\chi$ from various production mechanisms. 
The density parameter of $\chi$, $\Omega_\chi h^2$, is directly related to the energy density to entropy density ratio $\rho_\chi/s$ through the relation
\begin{align}
	\Omega_\chi h^2 \simeq 0.11 \left(\frac{\rho_\chi/s}{4\times 10^{-10}\,{\rm GeV}}\right).
\end{align}
The observed DM abundance is $\Omega_{\rm DM}h^2 \simeq 0.12$~\cite{Planck:2018vyg}.
Below we evaluate $\rho_\chi/s$ for each process.

\paragraph{Production from Higgs portal coupling}

The cross section is given by (\ref{sigmaHP}). The produced number density of $\chi$ particles in one Hubble time is given by
\begin{align}
	\Delta n_\chi \sim  n_\chi^{\rm (th) 2}  \left<\sigma v\right>_{\chi\chi\to \mathcal H \mathcal H}^{\rm (HP)} H^{-1}.
\end{align}
For $T \ll m_\chi$, the production rate is exponentially suppressed as $e^{-m_\chi/T}$. For $T \gtrsim m_\chi$, it is evaluated as
\begin{align}
	\Delta n_\chi \sim \frac{1}{64\pi^6}\sqrt{\frac{90}{g_*}}\times \begin{cases}
		\lambda^2 T_{\rm R}^2 M_{\rm Pl} & {\rm for}~T > T_{\rm R}\\
		\lambda^2 T^2 M_{\rm Pl} & {\rm for}~T < T_{\rm R}
	\end{cases},
\end{align}
where $g_*$ is the relativistic degrees of freedom, which we take $106.75$ in the case of our interest.
In both cases it decreases slower than $a^{-3}$ and hence the production at lower temperature is dominant as far as the Boltzmann suppression is not turned on. Therefore we need to evaluate it around $T\sim m_\chi$.
The resulting energy density to entropy density ratio is given by
\begin{align}
	\frac{\rho_\chi}{s} = m_\chi\frac{n_\chi}{s}\sim \frac{45}{128\pi^8 g_*}\sqrt{\frac{90}{g_*}}\times\begin{cases}
		\displaystyle \lambda^2 M_{\rm Pl}\left(\frac{T_{\rm R}}{m_\chi}\right)^7 & {\rm for}~T_{\rm R} < m_\chi < T_{\rm max} \\
		\displaystyle \lambda^2 M_{\rm Pl} & {\rm for}~m_\chi < T_{\rm R}
	\end{cases}.
	\label{rho_HP}
\end{align}
The overall numerical factor is about $3\times 10^{-7}$. For $m_\chi > T_{\rm max}$, the abundance is exponentially suppressed.
It is notable that the abundance is independent of $T_{\rm R}$ and $m_\chi$ if $m_\chi < T_{\rm R}$ and only depends on $\lambda$.
On the other hand, if $m_\chi > T_{\rm R}$, the abundance rapidly decreases as $(T_{\rm R}/m_\chi)^7$.
It is found that, for $\lambda \lesssim 10^{-11}$, it cannot account for the DM abundance irrespective of the choice of $m_\chi$ and $T_{\rm R}$.

\paragraph{Gravitational production from Standard Model particles}

The cross section is given by (\ref{sigmagrav_00}) or (\ref{sigmagrav_11}) and roughly expressed as $ \left<\sigma v\right>_{\chi\chi\to {\rm SM}\,{\rm SM}}^{\rm (grav)} = yT^2/M_{\rm Pl}^4$ with $y\sim 1$. Assuming $T \gtrsim m_\chi$, the produced number density per Hubble time is estimated as
\begin{align}
	\Delta n_\chi \sim  n_\chi^{\rm (th) 2}  \left<\sigma v\right>_{\chi\chi\to {\rm SM}\,{\rm SM}}^{\rm (grav)} H^{-1} 
	\sim \frac{1}{\pi^5}\sqrt{\frac{90}{g_*}}\times \begin{cases}
		\displaystyle \frac{T^4 T_{\rm R}^2}{M_{\rm Pl}^3} & {\rm for}~T > T_{\rm R}\\
		\displaystyle \frac{T^6}{M_{\rm Pl}^3}  & {\rm for}~T < T_{\rm R}
	\end{cases}.
\end{align}
For $T > T_{\rm R}$ $(T < T_{\rm R})$, the production at lower (higher) temperature is dominant. Thus we need to evaluate it at $T\sim m_\chi$ if $m_\chi > T_{\rm R}$, while at  $T\sim T_{\rm R}$ if $m_\chi < T_{\rm R}$. As a result, the abundance is given by
\begin{align}
	\frac{\rho_\chi}{s} = m_\chi\frac{n_\chi}{s}\sim 
	 \frac{45}{2\pi^7 g_*}\sqrt{\frac{90}{g_*}}\times\begin{cases}
		\displaystyle  \frac{T_{\rm R}^7}{m_\chi^3 M_{\rm Pl}^3} & {\rm for}~T_{\rm R} < m_\chi < T_{\rm max} \\
		\displaystyle  \frac{m_\chi T_{\rm R}^3}{M_{\rm Pl}^3} & {\rm for}~m_\chi < T_{\rm R}
	\end{cases}.
\end{align}
The overall numerical factor is $\sim 6\times 10^{-5}$.
Again, for $m_\chi > T_{\rm max}$, the abundance is exponentially suppressed.

\paragraph{Gravitational production from inflaton}

The production rate from the coherently oscillating inflaton is given by (\ref{Rgrav}). The produced number density per Hubble time is then estimated as $\Delta n_\chi \sim H^3$. Since it rapidly decreases faster than $a^{-3}$, the dominant contribution comes from earlier epoch, i.e., just after inflation. The resultant abundance is given by
\begin{align}
	\frac{\rho_\chi}{s} = \frac{3}{512\pi} \frac{m_\chi H_{\rm inf}T_{\rm R}}{M_{\rm Pl}^2}~~~{\rm for}~~m_\chi < m_\phi.
\end{align}
The abundance is exponentially suppressed for $m_\chi > m_\phi$.

\subsection{Dark matter abundance: numerical result}

In this section we show the results of the numerical solution to the Boltzmann equation (\ref{dotnchi}).

\begin{figure}
\begin{center}
   \includegraphics[width=8cm]{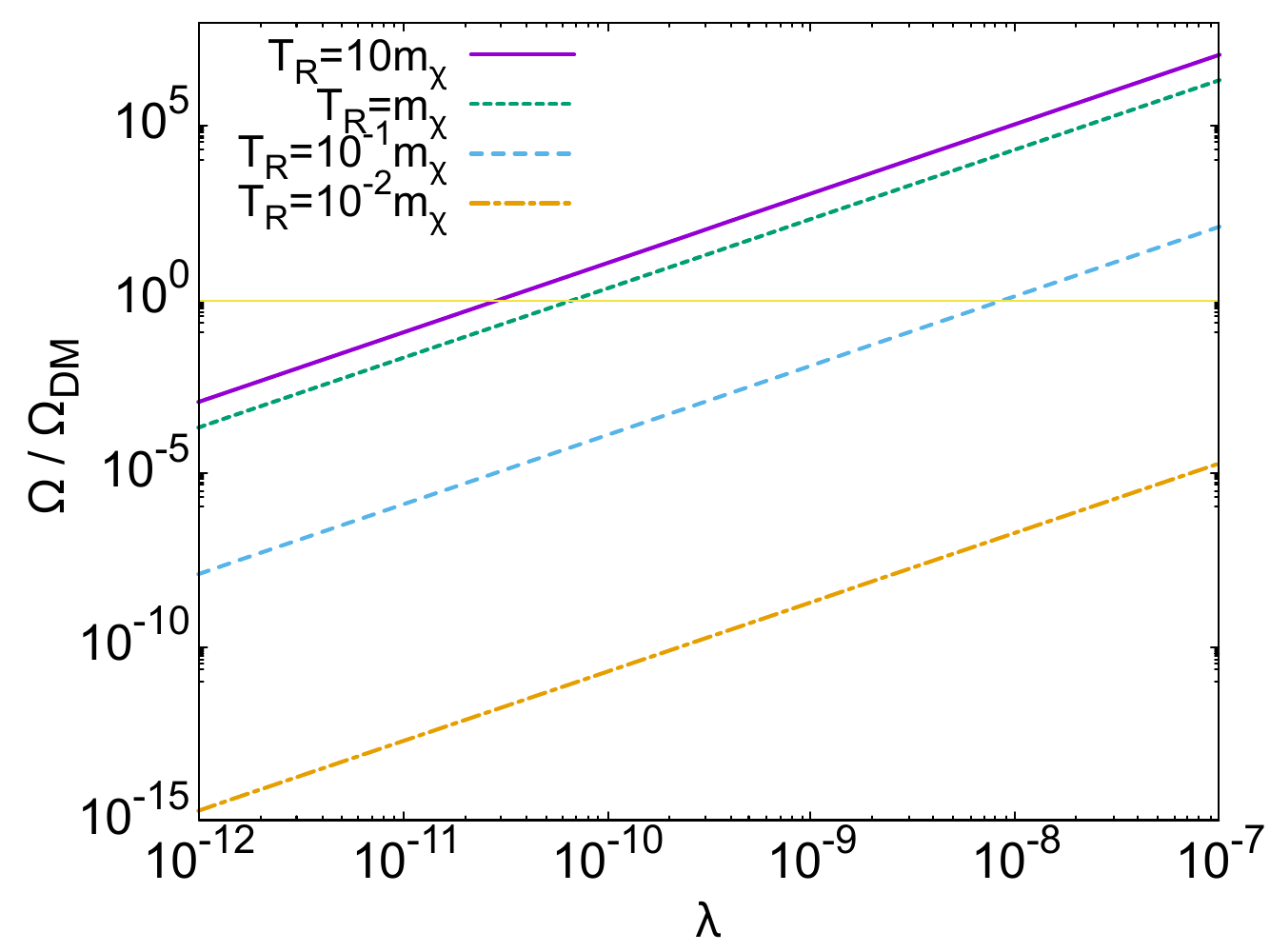}\includegraphics[width=8cm]{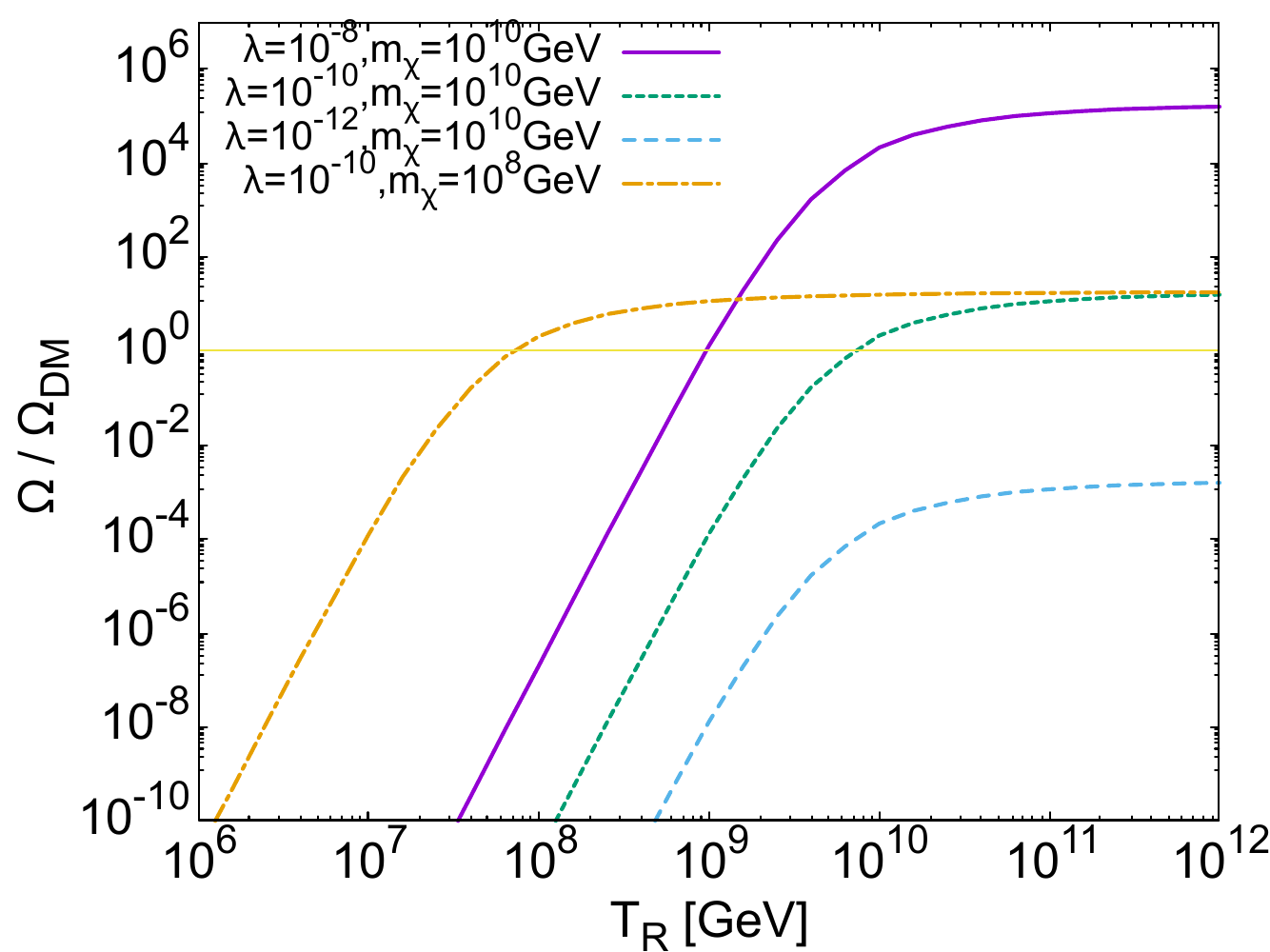}
  \end{center}
  \caption{The abundance of $\chi$, in terms of the density parameter $\Omega_\chi / \Omega_{\rm DM}$, originated only from the Higgs portal coupling. 
  (Left) The dependence on the Higgs portal coupling $\lambda$ for several choices of $T_{\rm R}/m_\chi$. (Right) The dependence on the reheating temperature $T_{\rm R}$ for several choices of $\lambda$ and $m_\chi$. }
  \label{fig:OHP}
\end{figure}

Fig.~\ref{fig:OHP} shows the results only including the Higgs portal coupling.
The abundance of $\chi$, in terms of the density parameter $\Omega_\chi / \Omega_{\rm DM}$, is shown as a function of $\lambda$ (left panel) and $T_{\rm R}$ (right panel).
One can clearly see the independence on $T_{\rm R}$ for $T_{\rm R} \gtrsim m_\chi$ and also the $T_{\rm R}^7$ dependence for $T_{\rm R} \ll m_\chi$, consistent with the analytic estimate (\ref{rho_HP}).

\begin{figure}
\begin{center}
   \includegraphics[width=8cm]{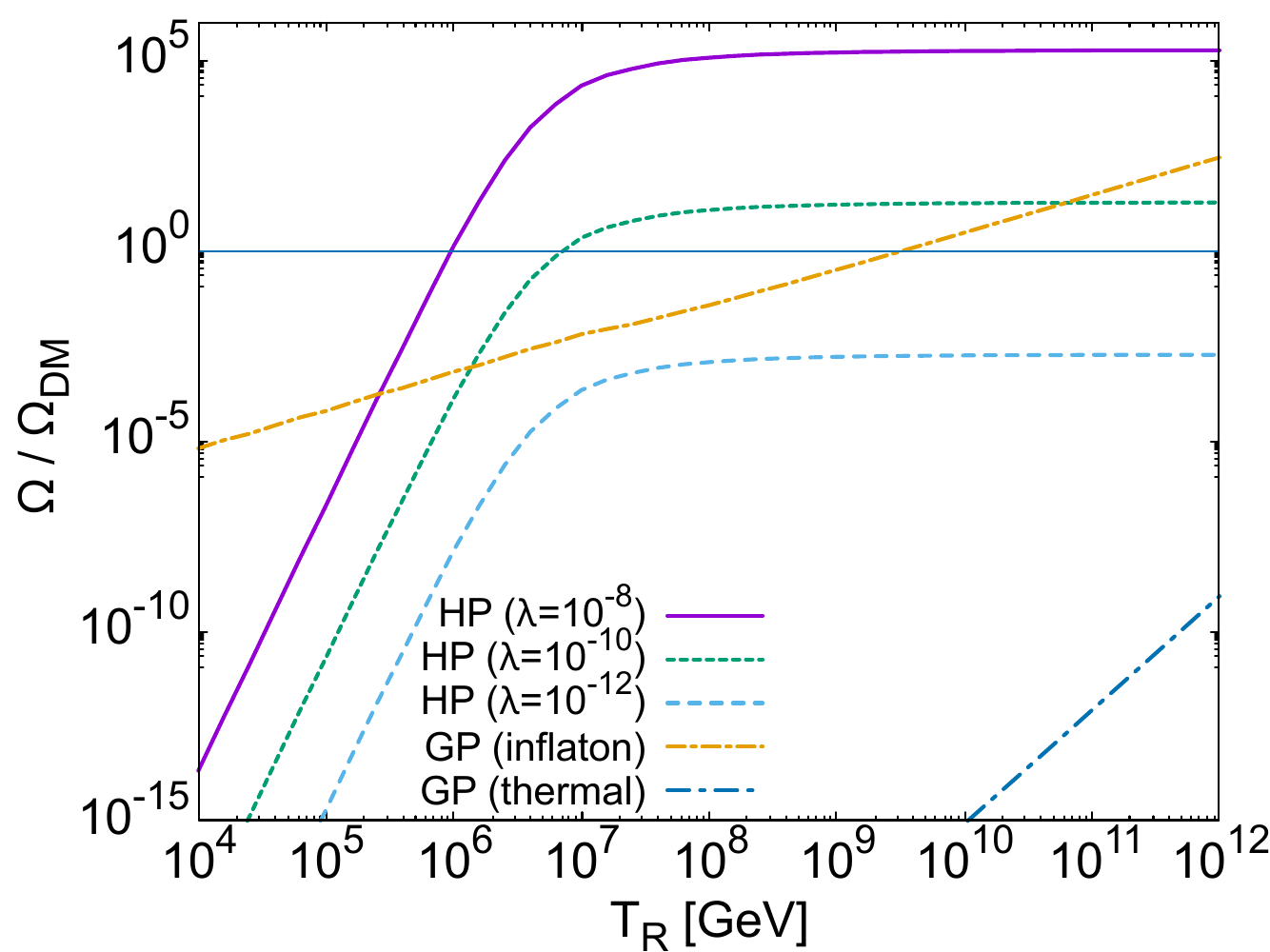}\includegraphics[width=8cm]{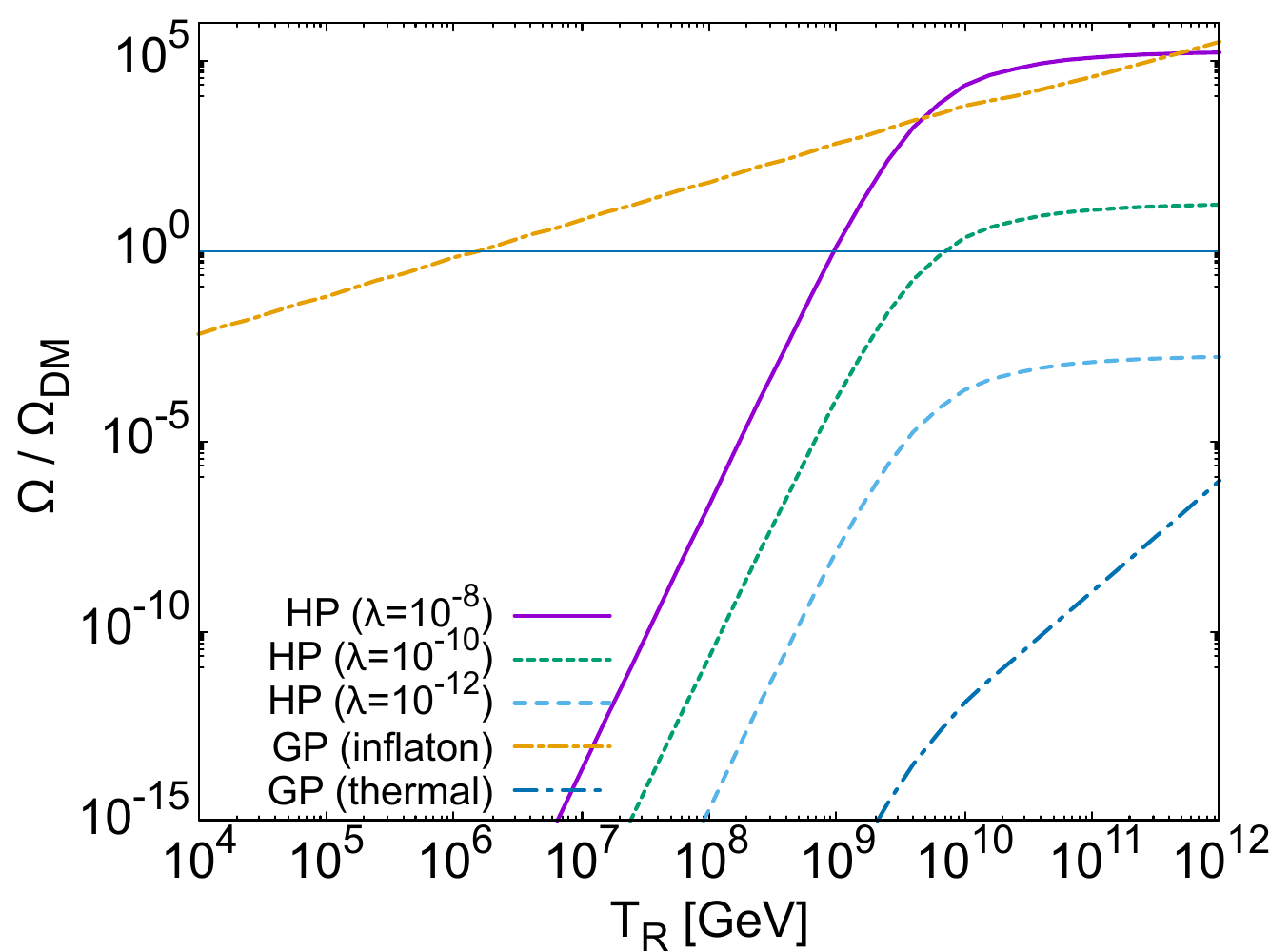}
  \end{center}
  \caption{The abundance of $\chi$, in terms of the density parameter $\Omega_\chi / \Omega_{\rm DM}$, from the Higgs portal coupling (``HP''), the gravitational production from thermal plasma (``GP (thermal)'') and the gravitational production from inflaton (``GP (inflaton)''). We have taken $m_\chi = 10^7\,{\rm GeV}$ ($10^{10}\,{\rm GeV}$) in the left (right) panel and $H_{\rm inf} = 10^{13}\,{\rm GeV}$. }
  \label{fig:mchi}
\end{figure}

Fig.~\ref{fig:mchi} shows the $\chi$ abundance from the Higgs portal coupling (``HP''), the gravitational production from thermal plasma (``GP (thermal)'') and the gravitational production from inflaton (``GP (inflaton)''). We have taken $m_\chi = 10^7\,{\rm GeV}$ ($10^{10}\,{\rm GeV}$) in the left (right) panel and $H_{\rm inf} = 10^{13}\,{\rm GeV}$.

\begin{figure}
\begin{center}
\includegraphics[width=16cm]{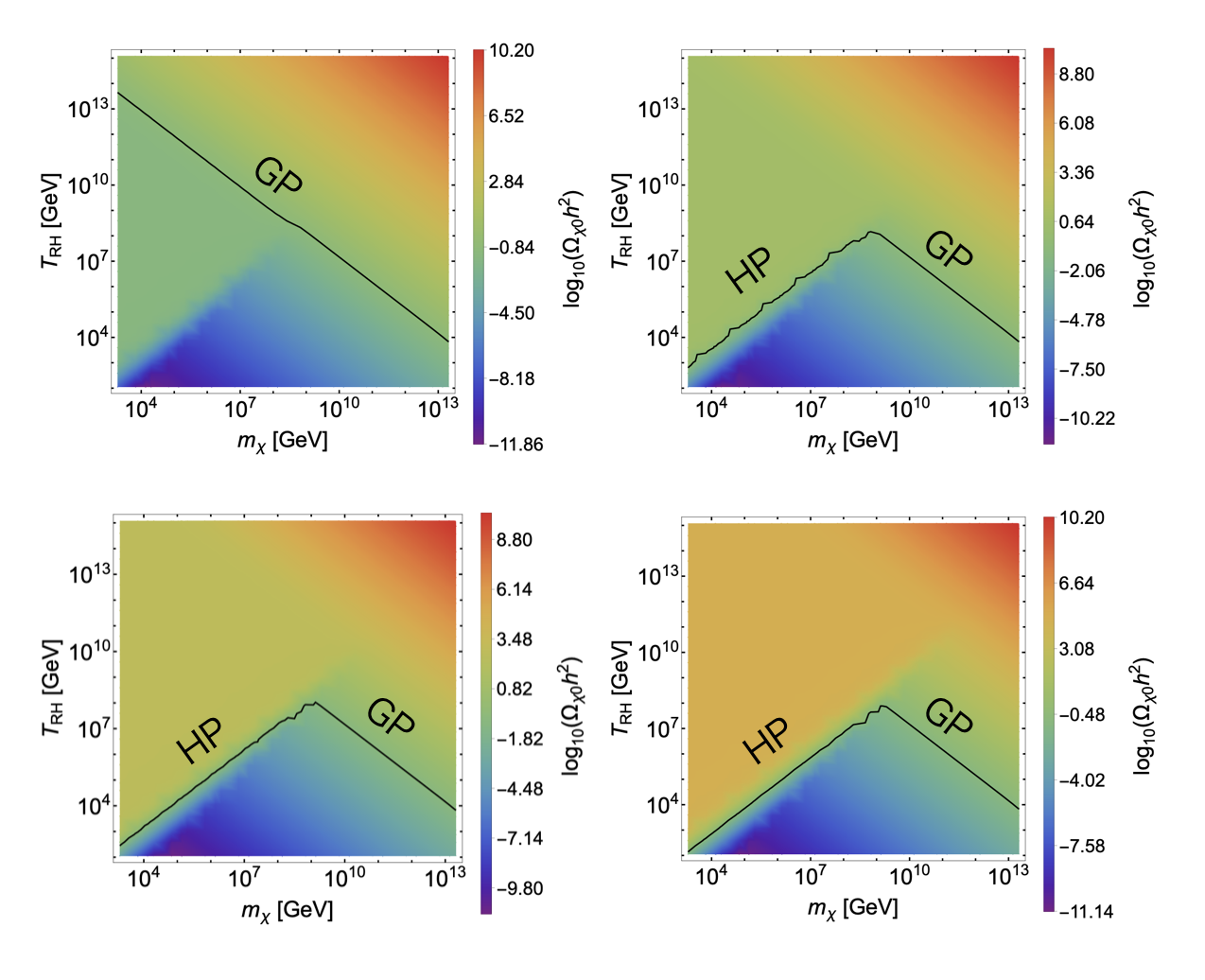}
  \end{center}
  \caption{The contours of DM abundance $\Omega_\chi h^2$ on the plane of the DM mass $m_\chi$ and the reheating temperature $T_{\rm R}$. We have taken the inflaton mass as $m_\phi = 3\times 10^{13}\,{\rm GeV}$ and the inflation Hubble scale $H_{\rm inf}=2\times 10^{13}\,{\rm GeV}$. The Higgs portal coupling $\lambda$ is taken to be $\lambda=1\times 10^{-11}$ (upper left), $\lambda=1\times 10^{-10}$ (upper right), $\lambda=1\times 10^{-9}$ (lower left), $\lambda=1\times 10^{-8}$ (lower right). The observed DM abundance is reproduced on the black line. The label ``GP'' and ``HP'' on the black line indicates that the corresponding part is dominated by the gravitational production and Higgs portal coupling, respectively.}
  \label{fig:dm}
\end{figure}

Fig.~\ref{fig:dm} shows the contours of DM abundance $\Omega_\chi h^2$ on the plane of the DM mass $m_\chi$ and the reheating temperature $T_{\rm R}$. We have taken the inflaton mass as $m_\phi = 3\times 10^{13}\,{\rm GeV}$ and the inflation Hubble scale $H_{\rm inf}=2\times 10^{13}\,{\rm GeV}$. 
These parameters for the inflaton sector corresponds to the so-called $\alpha$-attractor T model~\cite{Kallosh:2013hoa,Kallosh:2013yoa,Galante:2014ifa} (see App.~\ref{app:inf}).
The Higgs portal coupling $\lambda$ is taken to be $\lambda=1\times 10^{-11}$ (upper left), $\lambda=1\times 10^{-10}$ (upper right), $\lambda=1\times 10^{-9}$ (lower left), $\lambda=1\times 10^{-8}$ (lower right). The observed DM abundance is reproduced on the black line.
The label ``GP'' and ``HP'' on the black line indicates that the corresponding part is dominated by the gravitational production and Higgs portal coupling, respectively. Let us make several comments.

\begin{itemize}
\item In the upper left panel, the Higgs portal coupling is too small and it cannot account for the present DM abundance as understood from the analytic estimation (\ref{rho_HP}) and seen in the right panel of Fig.~\ref{fig:OHP}. Thus only the gravitational production from inflaton is important. 
\item For larger $\lambda$ $(\gtrsim 10^{-10})$, it is possible to account for the DM abundance through the Higgs portal coupling.
The positively sloped black lines in the upper right, lower left and lower right panels in Fig.~\ref{fig:dm} correspond to the production from the Higgs portal coupling.
\item The position of these lines on the $(m_\chi, T_{\rm R})$ plane do not much depend on $\lambda$ due to the strong $(T_{\rm R}/m_\chi)^7$ dependence on the final DM abundance (\ref{rho_HP}); they are slightly below the boundary  $T_{\rm R}=m_\chi$. 
As a result, the final result is not very sensitive to $\lambda$ for $\lambda \gtrsim 10^{-10}$.
\end{itemize}

\begin{figure}
\begin{center}
\includegraphics[width=16cm]{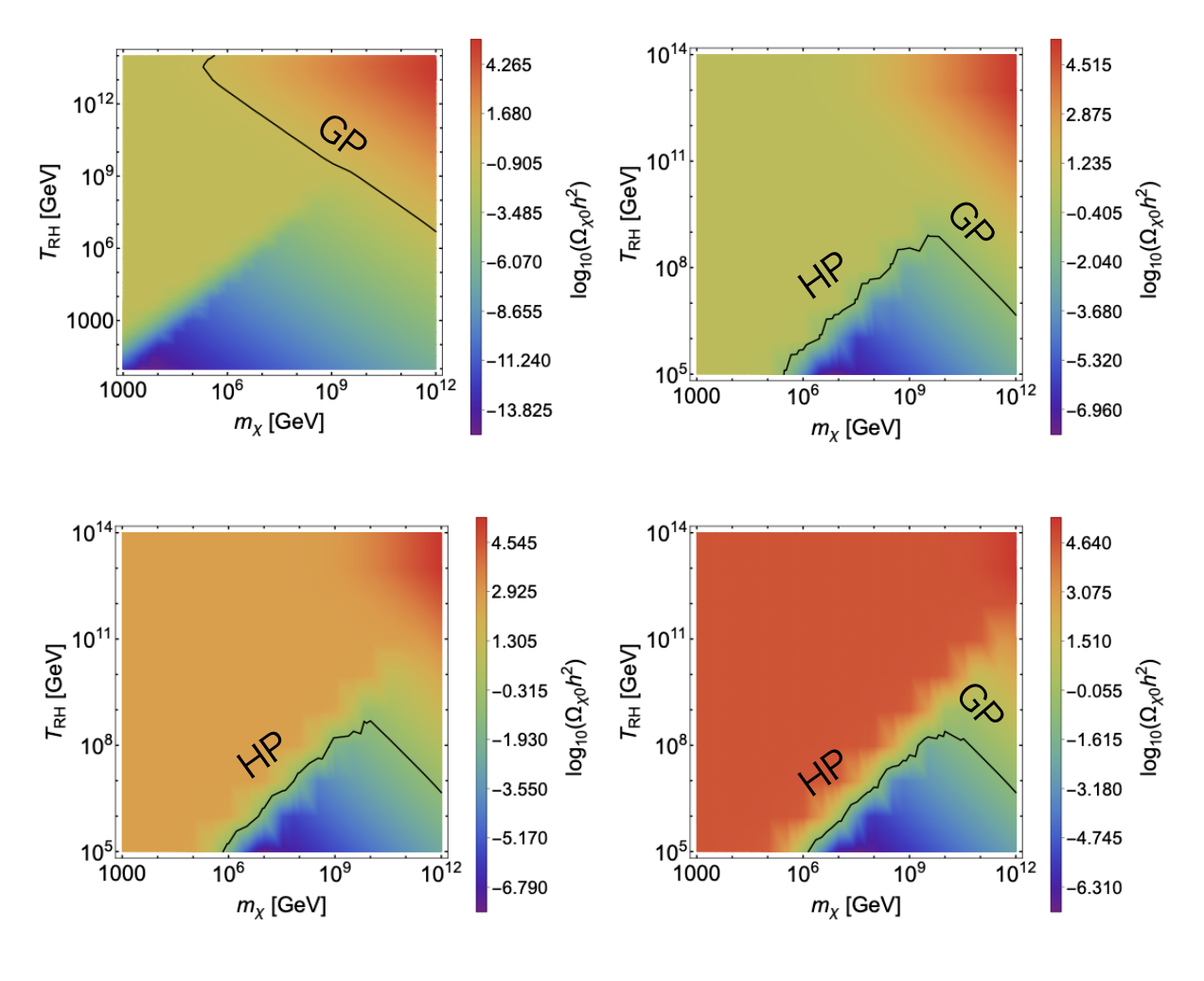}
  \end{center}
  \caption{The same as Fig.~\ref{fig:dm}, but for $m_\phi = 2\times 10^{12}\,{\rm GeV}$ and $H_{\rm inf}=1\times 10^{10}\,{\rm GeV}$.}
 \label{fig:dm2}
\end{figure}

Fig.~\ref{fig:dm2} shows the case for $m_\phi = 2\times 10^{12}\,{\rm GeV}$ and $H_{\rm inf}=1\times 10^{10}\,{\rm GeV}$, as typical parameters for new inflation (see App.~\ref{app:inf}). It is notable that there is an upper bound on the reheating temperature as $T_{\rm R} \lesssim 10^9\,{\rm GeV}$ for $\lambda > 10^{-10}$.
Due to lower inflation scale the gravitational contribution is smaller and higher $T_{\rm R}$ is required for reproducing the DM abundance at high mass range represented by negatively sloped lines. On the other hand, it does not much affect the Higgs-portal contribution as represented by positively sloped lines.

\section{Other production mechanisms} \label{sec:other}

In this section we consider several other DM production mechanisms, which may naturally coexist in our scenario.

\subsection{Misalignment production} \label{sec:mis}

We have neglected the misalignment production of the scalar DM.
In general, if $m_\chi \lesssim H_{\rm inf}$, the $\chi$ field can have a homogeneous field value during inflation $\chi(\vec x)=\chi_0$, which in turn becomes coherent oscillation with an initial amplitude $\chi_0$ when $H \sim m_\chi$: this is called as the misalignment production. The abundance of $\chi$ coherent oscillation is given by
\begin{align}
	\frac{\rho_\chi}{s} = \frac{1}{8}\left(\frac{\chi_0}{M_{\rm Pl}}\right)^2 \times {\rm min}\left[ T_{\rm R},T_{\rm osc}\right],
\end{align}
where $T_{\rm osc}=(90/\pi^2 g_*)^{1/4}\sqrt{m_\chi M_{\rm Pl}}$. 
The natural value of $\chi_0$ is highly model dependent. If the $\chi$ potential is quadratic and inflation lasts long enough, the long wave quantum fluctuation of $\chi$ accumulates to the asymptotic value $\chi_0 \sim H_{\rm inf}^2 / m_\chi$~\cite{Linde:1990flp}. 
If $\chi$ has an additional potential term like $V(\chi) \sim \lambda_\chi \chi^4$, a natural value might be $\chi_0\sim H_{\rm inf}/\lambda_\chi^{1/4}$. In these cases, the misalignment contribution is often too large to be consistent with the observed DM abundance.
However, the misalignment production is easily suppressed if there is a coupling between $\chi$ and Ricci scalar as
\begin{align}
	\mathcal L = \frac{\xi}{2}R \chi^2,
	\label{Rchi2}
\end{align}
where $\xi$ is a parameter of order unity. This gives the (positive) mass of the order of the Hubble scale during inflation and $\chi$ dynamically relaxes to the minimum $\chi=0$ during inflation, and hence there is no misalignment production.
Note that the term (\ref{Rchi2}) itself affects the gravitational production from the inflaton coherent oscillation. The production rate (\ref{Rgrav}) should be multiplied by the overall factor $(1-6\xi)^2$. Unless the coupling is very close to the conformal value $(\xi=1/6)$, the DM abundance calculated so far is still reasonable as an order of magnitude estimation.

\subsection{Inflaton portal coupling}

One may add a following portal coupling between the inflaton and DM in the Lagrangian like,
\begin{align}
	\mathcal L = -\frac{1}{2}\lambda_{\phi\chi}\phi^2\chi^2.
	\label{inflaton-portal}
\end{align}
Let us suppose that the inflaton is oscillating around $\phi=0$.
In order for this coupling not to affect the inflaton potential through radiative correction, we obtain a constraint as $\lambda_{\phi\chi} \lesssim 10^{-6}$.
Still such a coupling may significantly affect the DM abundance through particle production from inflaton oscillation~\cite{Dolgov:1989us,Traschen:1990sw,Kofman:1994rk,Shtanov:1994ce,Kofman:1997yn}.
However, such a coupling is highly dependent on the inflaton properties.
The most successful inflation models require some tuning or small parameters for the inflaton potential itself, indicating that the inflaton sector may be more or less separated from the other sector, although still small coupling is required for successful reheating.
For example, one may consider the inflaton mass term as an explicit breaking of the shift symmetry $\phi\to\phi + {\rm const}$.\footnote{
	Note that the coupling like $\mathcal L \sim \phi F_{\mu\nu}\tilde F^{\mu\nu}$, where $F_{\mu\nu}$ is the gauge boson field strength and $\tilde F^{\mu\nu}$ is its dual, is allowed by the shift symmetry. Such an axionic coupling may be responsible for the successful reheating.
} In this sense, the measure of the shift symmetry breaking may be given by $m_\phi^2 / M_{\rm Pl}^2 $ and hence we may naturally have a small inflaton portal coupling $\lambda_{\phi\chi} \lesssim m_\phi^2/M_{\rm Pl}^2$.
This indicates that the particle production is not in the broad resonance regime. Thus the ratio between the production rate due to the inflaton portal coupling and the gravitational coupling is given by
\begin{align}
	\frac{R^{\rm (portal)}_{\phi\phi\to\chi\chi}}{R^{\rm (grav)}_{\phi\phi\to\chi\chi}} \sim \frac{\lambda_{\phi\chi}^2 M_{\rm Pl}^4}{m_\phi^4}.
\end{align}
It is at most $\mathcal O(1)$ and hence the estimate of the DM abundance in the previous section is not much affected.
In addition to the shift symmetry of $\phi$, the property of $\chi$ may also cause extra suppression factor for the coupling $\lambda_{\phi\chi}$, by noting that the $\chi$ potential may be even flatter than the inflaton.\footnote{
	If the inflaton oscillates around a finite vacuum expectation value, as is the case for new inflation for example, the portal coupling induces the inflaton decay $\phi\to\chi\chi$. In this case, the branching ratio of the inflaton decay to $\chi$ must be extremely suppressed in order to avoid the DM overproduction. In such a case, this extra suppression may be helpful.
}
Again, however, we stress that this argument on the smallness of $\lambda_{\phi\chi}$ is not rigorously justified and it is possible that the inflaton portal coupling has a more significant impact on the DM abundance.

\section{Conclusions and discussion} \label{sec:con}

We have studied the production process of one of the simplest DM model, a scalar particle $\chi$ with Higgs portal coupling. It is known that consistent abundance of superheavy DM is obtained for a (very) small Higgs portal coupling $\lambda$ through the freeze-in production.
For such a small coupling, however, the gravitational production process can also be important.
One process is the annihilation of Standard Model particles in thermal bath into $\chi$ pair through the s-channel gravition exchange. Another process is caused by the inflaton coherent oscillation.
We take account of all of these processes and evaluated the final DM abundance for various parameters.
The model parameters are the DM mass $m_\chi$, Higgs portal coupling $\lambda$, reheating temperature $T_{\rm R}$ and Hubble scale at the end of inflation $H_{\rm inf}$.
The results are shown in Fig.~\ref{fig:dm}. 
We find that the Higgs portal coupling cannot account for the present DM abundance for $\lambda \lesssim 10^{-11}$ no matter how other parameters are chosen, as understood from the analytical estimation (\ref{rho_HP}).
The gravitational production, on the other hand, can account for the present DM abundance for wide parameter ranges by choosing $T_{\rm R}$ and/or $H_{\rm inf}$ appropriately.
In particular, the inflaton-induced gravitational production is often dominant compared with that from the Standard Model particles.
For larger $\lambda$, the Higgs portal production can dominate over the gravitational one, but still the gravitational production is important for heavier DM as seen in Figs.~\ref{fig:dm} and \ref{fig:dm2}. 
Importantly, if the Higgs portal contribution is dominant, the parameter region of $(m_\chi, T_{\rm R})$ required for consistent DM abundance does not much depend on $\lambda$, due to the relatively large dependence of $\Omega_\chi$ on these parameters: $\Omega_\chi\propto \lambda^2 (T_{\rm R}/m_\chi)^7$ (\ref{rho_HP}).

The final abundance depends on the inflationary parameters $T_{\rm R}$ and $H_{\rm inf}$. The inflationary scale $H_{\rm inf}$ may be probed by future observations of the B-mode polarization in the cosmic microwave background, e,g., with the LiteBIRD satellite~\cite{LiteBIRD:2022cnt} or CMB-S4~\cite{Abazajian:2019eic}.
The reheating temperature $T_{\rm R}$ may also be probed by the direct observation of primordial gravitational waves by the space laser interferometers~\cite{Nakayama:2008ip,Nakayama:2008wy,Kuroyanagi:2008ye,Kuroyanagi:2011fy,Kuroyanagi:2014qza}, such as DECIGO~\cite{Kawamura:2020pcg}.
If $T_{\rm R}$ turns out to be higher than $T_{\rm R} \gtrsim 10^{10}\,{\rm GeV}$, the production through the Higgs portal coupling may be excluded, while the gravitational production still may work for appropriate DM mass $m_\chi$. It will be a crucial information for DM model.  
This value of the reheating temperature is also interesting from the viewpoint of thermal leptogenesis~\cite{Fukugita:1986hr}, since successful generation of baryon asymmetry of the universe requires $T_{\rm R} \gtrsim 10^9$\,GeV~\cite{Giudice:2003jh,Buchmuller:2004nz}.
We again emphasize that a nontrivial information on $T_{\rm R}$ is the outcome of the combination of the Higgs portal production and gravitational production of DM.
Since the gravitational production is ubiquitous, we must always be careful about the gravitational contribution to the DM abundance, especially if the DM coupling is very weak.

\section*{Acknowledgment}

This work was supported by World Premier International Research Center Initiative (WPI), MEXT, Japan.

\appendix
\section{Cross section} \label{app:th}

\subsection{Thermally averaged cross section}

Thermally averaged cross section for the process $12\to 34$ is given by
\begin{align}
	\left<\sigma v\right>_{12\to 34} = \frac{1}{n_1 n_2} & \int \frac{g_1 d^3p_1}{(2\pi)^3 2E_1}\frac{g_2 d^3p_2}{(2\pi)^3 2E_2} \frac{g_3 d^3p_3}{(2\pi)^3 2E_3}\frac{g_4 d^3p_4}{(2\pi)^3 2E_4} \nonumber \\
	&\times  (2\pi)^4\delta(p_1+p_2-p_3-p_4) \left|\mathcal M_{12\to 34}\right|^2 f_1 f_2(1\pm f_3) (1\pm f_4),
\end{align}
where $g_i$ is the internal degrees of freedom for a particle $i$ ($g_i=1$ for a scalar, $g_i=2$ for a fermion, etc),\footnote{
	To avoid the overcounting, $\left|\mathcal M_{12\to 34}\right|^2$ should be interpreted as the square amplitude averaged over the final spin or polarization states.
} and $f_i$ is the distribution function for the particle species $i$ and the plus (minus) sign corresponds to the boson (fermion).
With the Boltzmann approximation $f_i \simeq e^{-E_i/T}$, the number density is given by
\begin{align}
	n_i^{\rm (th)} = g_i \frac{m_i^2 T}{2\pi^2} K_2\left(\frac{m_i}{T}\right),  \label{ni}
\end{align}
if species 1 and 2 are thermalized, where $K_2$ is the modified Bessel function of the second kind of order 2.\footnote{
		The high temperature limit of (\ref{ni}) slightly differs from the full relativistic result by a factor $\zeta(3) \simeq 1.202$ for a boson and $\zeta(3)\times 3/4 \simeq 0.9015$ for a fermion.
}
With the Boltzmann approximation for all particles, we find a relation $\left<\sigma v\right>_{12\to 34} n_1^{\rm (th)}n_2^{\rm (th)} = \left<\sigma v\right>_{34\to 12} n_3^{\rm (th)}n_4^{\rm (th)}$ where $n_3$ and $n_4$ are hypothetical thermal number density even if they are not thermalized actually.
Below we mainly focus on the case where the ``initial state'' 3 and 4 are $\chi$ particle, which are not thermalized actually.

Following Ref.~\cite{Gondolo:1990dk}, within the Boltzmann approximation and when the initial states have the same mass $m$, the cross section is given by
\begin{align}
	\left<\sigma v\right>= \frac{1}{8m^4 T K_2^2(m/T)}\int_{4m^2}^\infty ds (s-4m^2)\sqrt{s}K_1\left(\frac{\sqrt s}{T}\right) \sigma(s),
	\label{sigmav}
\end{align}
where $K_n$ is the modified Bessel function of the second kind of order $n$ and
\begin{align}
	\sigma(s) &= \frac{1}{2\sqrt{s(s-4m^2)}} \int \frac{g_3 d^3p_3}{(2\pi)^3 2E_3}\frac{g_4 d^3p_4}{(2\pi)^3 2E_4}(2\pi)^4\delta^4(p_1+p_2-p_3-p_4)\left|\mathcal M_{12\to 34}\right|^2 \nonumber\\
	&= \frac{g_3 g_4}{16\pi s(s-4m^2)}\int_{t_-}^{t_+}dt \left|\mathcal M_{12\to 34}\right|^2,
\end{align}
where $t_{\pm} \equiv m^2-s/2 \pm \sqrt{s(s-4m^2)}/2$. Here $s=(p_1+p_2)^2$ and $t=(p_1-p_3)^2$ are the Mandelstam variables.
We neglected the mass of particles 3 and 4 in the second line, which is justified for our purpose since the final state particles are the Standard Model particles and they are much lighter than $\chi$.
Note that the symmetry factor $1/2$ should be multiplied when calculating $\sigma(s)$ if the final states are identical particles.\footnote{
	The symmetry factor for identical $\chi$ particles as an initial state are already cancelled in the Boltzmann equation (\ref{dotnchi}) by the fact that two $\chi$ particles are created in one annihilation process. 
}

\paragraph{Production from Higgs portal coupling}

For the process $\chi\chi\to \mathcal H \mathcal H$ through the Higgs portal coupling, the amplitude is $|\mathcal M|^2=\lambda^2$. Taking account of four real degrees of freedom of $\mathcal H$, $\sigma(s)$ is given by
\begin{align}
	\sigma(s)^{\rm (HP)}_{\chi\chi\to \mathcal H \mathcal H} = \frac{\lambda^2}{16\pi} \frac{1}{\sqrt{s(s-4m_\chi^2)}}.
\end{align}
To derive thermally averaged cross section (\ref{sigmav}), we need to calculate the integral involving the Bessel function. In the next subsection we provide a useful formula for such an integral.

\paragraph{Gravitational production from Standard Model particles}

For the annihilation processes into the Standard Model particles into spin-$0, \frac{1}{2}, 1$ particles mediated by the s-channel graviton, we obtain the spin- or polarization-averaged square amplitude as~\cite{Garny:2017kha,Clery:2021bwz}
\begin{align}
	&\left|\mathcal M_{\chi\chi\to 00}\right|^2 = \frac{(t-m_\chi^2)^2(s+t-m_\chi^2)^2}{M_{\rm Pl}^4 s^2}, \\
	&\left|\mathcal M_{\chi\chi\to \frac{1}{2}\frac{1}{2}}\right|^2 = -\frac{(s+2t-2m_\chi^2)^2(t(s+t)-2m_\chi^2 t +m_\chi^4)}{16M_{\rm Pl}^4 s^2}, \\
	&\left|\mathcal M_{\chi\chi\to 11}\right|^2 = \frac{(t(s+t)-2m_\chi^2t + m_\chi^4)^2}{2M_{\rm Pl}^4 s^2},
\end{align}
where we have neglected the mass of Standard Model particles.
Note that we assumed a chiral fermion for a spin $\frac{1}{2}$ particle. They lead to $\sigma(s)$ as
\begin{align}
	&\sigma(s)^{\rm (grav)}_{\chi\chi\to 00} = \frac{1}{960\pi M_{\rm Pl}^4} \frac{6m_\chi^4 + 2m_\chi^2 s + s^2}{\sqrt{s(s-4m_\chi^2)}},\\
	&4\sigma(s)^{\rm (grav)}_{\chi\chi\to \frac{1}{2}\frac{1}{2}}=\sigma(s)^{\rm (grav)}_{\chi\chi\to 11} = \frac{1}{480\pi M_{\rm Pl}^4} \frac{(s-4m_\chi^2)^{3/2}}{\sqrt{s}}.
\end{align}
We have multiplied a symmetry factor $1/2$ for spin 0 and 1.
For deriving thermally averaged cross section, we need to evaluate the integral (\ref{sigmav}). We present some useful integral formulae in the next subsection. The results for thermally averaged cross section is given by (\ref{sigmagrav_00}) and (\ref{sigmagrav_11}), respectively.

\subsection{Useful formula}

When calculating the thermally averaged cross section, we encounter the integral of the form
\begin{align}
	F_n(\beta) \equiv \int_{4m^2}^\infty ds \sqrt{s-4m^2} s^n K_1(\sqrt{s}\beta),
	\label{Fn}
\end{align}
where $\beta\equiv 1/T$. Note that the case of $n=0$ is explicitly calculated as
\begin{align}
	F_0(\beta) = \frac{4m^2 K_1^2(m\beta)}{\beta}.
\end{align}
For later convenience, we also define
\begin{align}
	G_{n+\frac{1}{2}}(\beta) \equiv \int_{4m^2}^\infty ds \sqrt{s-4m^2} s^{n+1/2} K_2(\sqrt{s}\beta).
	\label{Gn}
\end{align}
In Refs.~\cite{Kolb:2017jvz,Aoki:2022dzd} they are evaluated in terms of the Meijer G-function. Actually, however, it can be expressed in terms of the Bessel function in a simpler way~\cite{Garny:2017kha}. Here is the procedure.
By differentiate (\ref{Fn}) with respect to $\beta$, we find\footnote{
	We use the derivative formula for the Bessel function: $K_n'(x)=-K_{n-1}(x)-(n/x)K_n(x) =-K_{n+1}(x)+(n/x)K_n(x)$.
}
\begin{align}
	\frac{F_n(\beta)}{\beta} - \frac{\partial F_n(\beta)}{\partial\beta} = G_{n+\frac{1}{2}}(\beta).
\end{align}
By differentiate (\ref{Gn}) with respect to $\beta$, we find
\begin{align}
	\frac{2G_{n+\frac{1}{2}}(\beta)}{\beta} + \frac{\partial G_{n+\frac{1}{2}}(\beta)}{\partial\beta} = -F_{n+1}(\beta).
\end{align}
Using these relations, we can recursively derive $F_n(\beta)$ for $n\geq 0$ in terms of the Bessel function.
Here we give explicit results, which will be needed to calculate the thermally averaged cross section:
\begin{align}
	&G_{\frac{1}{2}}(\beta) = \frac{8m^3 K_1(m\beta) K_2(m\beta)}{\beta},\\
	&F_1(\beta) =  \frac{8m^4\left[ \left( K_1(m\beta) \right)^2 + \left( K_2(m\beta) \right)^2 \right]}{\beta},\\
	&G_{\frac{3}{2}}(\beta) = \frac{16m^4K_2(m\beta)\left[ 2m\beta K_1(m\beta) +3 K_2(m\beta) \right]}{\beta^2},\\
	&F_2(\beta) =  \frac{32m^4\left[ m^2\beta^2 (K_1(m\beta))^2 + 3m\beta K_1(m\beta) K_2(m\beta) +(6+m^2\beta^2) (K_2(m\beta))^2 \right]}{\beta^3}.
\end{align}

\section{Inflation models} \label{app:inf}

For calculating the gravitational production rate concretely, we must specify the inflaton mass $m_\phi$ and the Hubble scale at the end of inflation $H_{\rm inf}$. These parameters significantly depend on inflation models. 
In this Appendix we show several inflation models that are consistent with the Planck result.

\subsection{$\alpha$-attractor T model}

Let us consider the inflation model given by the Lagrangian~\cite{Kallosh:2013hoa,Kallosh:2013yoa,Galante:2014ifa}
\begin{align}
	\mathcal L = \frac{\frac{1}{2}(\partial \phi)^2}{\left(1-\frac{\phi^2}{\Lambda^2}\right)^2} - \lambda_\phi M^{4-n} \phi^n.
\end{align}
In terms of the canonically normalized inflaton $\varphi$, which is given by the relation
\begin{align}
	\phi = \Lambda \tanh\left(\frac{\varphi}{\Lambda}\right),
\end{align}
the Lagrangian is rewritten as
\begin{align}
	\mathcal L = \frac{1}{2}(\partial \varphi)^2 - \lambda_\phi M^{4-n} \Lambda^n \tanh^n\left(\frac{\varphi}{\Lambda}\right).
\end{align}
Thus the potential for the canonical inflaton $\varphi$ is flat for $|\varphi| \gg \Lambda$. The inflaton dynamics is solved under the slow-roll approximation as
\begin{align}
	\varphi(N) \simeq \Lambda \log\left(\frac{\sqrt{8nN} M_{\rm Pl}}{\Lambda}\right),
\end{align}
with $N$ being the e-folding number at the horizon crossing of the present cosmological scales. The scalar spectral index $n_s$ and the tensor-to-scalar ratio $r$ are calculated in the standard manner~\cite{Baumann:2009ds} as
\begin{align}
	n_s=1-\frac{2}{N},~~~r= \frac{2}{N^2}\left(\frac{\Lambda}{M_{\rm Pl}}\right)^2.
\end{align}
For $\Lambda\lesssim M_{\rm Pl}$ both $n_s$ and $r$ are consistent with observations. The power spectrum of the curvature perturbation is given by
\begin{align}
	\mathcal P_\zeta = \frac{N^2 \lambda_\phi M^{4-n} \Lambda^{n-2}}{3\pi^2 M_{\rm Pl}^2}.
\end{align}
This should be $\mathcal P_\zeta \simeq 2.1\times 10^{-9}$~\cite{Planck:2018jri}. For $n=2$, it is independent of $\Lambda$ and the inflaton mass around the minimum $\varphi=0$ is also independent of $\Lambda$ as
\begin{align}
	m_\phi = \sqrt{2\lambda_\phi} M = \frac{\sqrt{6\pi^2\mathcal P_\zeta} M_{\rm Pl}}{N} \simeq 1.4\times 10^{13}\,{\rm GeV}\left(\frac{60}{N}\right).
\end{align}
The Hubble scale at the end of inflation depends on $\Lambda$ as
\begin{align}
	H_{\rm inf} = \frac{m_\phi \Lambda}{\sqrt{6} M_{\rm Pl}} \simeq 6\times 10^{12}\,{\rm GeV}\left(\frac{\Lambda}{M_{\rm Pl}}\right)\left(\frac{60}{N}\right).
\end{align}

\subsection{New inflation}

Let us consider a new inflation model given by the Lagrangian~\cite{Kumekawa:1994gx,Izawa:1996dv,Asaka:1999jb,Senoguz:2004ky,Kohri:2007gq,Nakayama:2012dw,Ema:2017rkk}
\begin{align}
	\mathcal L = \frac{1}{2}(\partial\phi)^2 - \Lambda^4\left[ 1-\left(\frac{\phi}{v}\right)^n \right]^2.
\end{align}
Following the standard procedure for slow-roll inflation, the scalar spectral index and the tensor-to-scalar ratio are given by
\begin{align}
	n_s = 1- \frac{2}{N}\frac{n-1}{n-2},~~~~~~r=\frac{16n}{N(n-2)}\left(\frac{1}{2Nn(n-2)} \frac{v^2}{M_{\rm Pl}^2}\right)^{n/(n-2)}.
\end{align}
For $n\geq 6$ we can obtain the spectral index consistent with the observation. The power spectrum of the curvature perturbation is given by
\begin{align}
	\mathcal P_\zeta = \frac{\left[2n((n-2)N)^{n-1}\right]^{2/(n-2)}}{12\pi^2} \frac{\Lambda^4}{(v^n M^{n-4}_{\rm Pl})^{2/(n-2)}}
\end{align}
By requiring $\mathcal P_\zeta \simeq 2.1\times 10^{-9}$, the inflaton mass for $n=6$ is given by
\begin{align}
	m_\phi = \frac{\sqrt{2} n \Lambda^2}{v} \simeq \sqrt{72 \mathcal P_\zeta}\left(\frac{9\pi^4}{1024 N^5}\right)^{1/4}\sqrt{v M_{\rm Pl}}
	\simeq 2\times 10^{12}\,{\rm GeV}\left(\frac{v/M_{\rm Pl}}{10^{-1}}\right)^{1/2}.
\end{align}
The inflation Hubble scale is given by
\begin{align}
	H_{\rm inf} \simeq \frac{\Lambda^2}{\sqrt{3} M_{\rm Pl}} \simeq 1\times 10^{10}\,{\rm GeV} \left(\frac{v/M_{\rm Pl}}{10^{-1}}\right)^{3/2}.
\end{align}

\subsection{Starobinsky inflation}

Let us also mention the Starobinsky inflation~\cite{Starobinsky:1980te}. Instead of (\ref{action}), the action we consider is
\begin{align}
	S = \int d^4x \sqrt{-g_{\rm J}}\left( -\frac{M_{\rm Pl}^2}{2}R_{\rm J} + \frac{R_{\rm J}^2}{12\mu^2} + \mathcal L_{\chi,{\rm J}} + \mathcal L_{\rm HP,{\rm J}} +  \mathcal L_{\rm SM, J} \right),
\end{align}
where the subscript J indicates the Jordan frame, and
\begin{align}
	 \mathcal L_{\chi,{\rm J}} = \frac{1}{2}g_{\rm J}^{\mu\nu}\partial_\mu \chi \partial_\nu \chi - \frac{1}{2}m_\chi^2\chi^2 + \frac{\xi}{2} R_{\rm J}\chi^2.
\end{align}
There appears a dynamical degree of freedom in the gravity sector due to the $R_{\rm J}^2$ term. This is evident after the conformal transformation, 
\begin{align}
	g_{\mu\nu} = \exp\left( \sqrt{\frac{2}{3}}\frac{\phi}{M_{\rm Pl}}\right) g_{\mu\nu}^{\rm J}.
\end{align}
With this transformation, we can go to the Einstein frame in which the gravity sector becomes the Einstein-Hilbert action and the scalaron field $\phi$ is canonically normalized. The resulting action is
\begin{align}
	S = \int d^4x \sqrt{-g}\left( -\frac{M_{\rm Pl}^2}{2}R + \mathcal L_\phi + \mathcal L_{\chi} + \mathcal L_{\rm HP} +  \mathcal L_{\rm SM} \right),
\end{align}
where
\begin{align}
	\mathcal L_\phi = \frac{1}{2}g^{\mu\nu}\partial_\mu \phi \partial_\nu\phi - \frac{3\mu^2 M_{\rm Pl}^2}{4}\left[1-\exp \left(-\sqrt{\frac{2}{3}}\frac{\phi}{M_{\rm Pl}}\right)\right]^2,
\end{align}
and $\mathcal L_{\chi}$ and $\mathcal L_{\rm SM}$ include the $\phi$ interaction with each sector.
With this inflaton potential, the inflation happens for $\phi \gg M_{\rm Pl}$, and the scalar spectral index, the tensor-to-scalar ratio and the power spectrum of the curvature perturbation are given by
\begin{align}
	n_s = 1 - \frac{2}{N},~~~ r=\frac{12}{N^2},~~~\mathcal P_\zeta = \frac{\mu^2 N^2}{24\pi^2 M_{\rm Pl}^2}.
\end{align}
Thus $n_s$ and $r$ are consistent with the current observations. From the condition $\mathcal P_\zeta\simeq 2.1\times 10^{-9}$, the inflaton mass is determined as $m_\phi=\mu \simeq 3\times 10^{13}\,{\rm GeV}$.
The Hubble scale at the end of inflation is given by $H_{\rm inf} \simeq \mu/2 \sim 2\times 10^{13}\,{\rm GeV}$.

Note that in this model the inflaton decays into $\chi$ pair as $\phi\to\chi\chi$, which overproduces the $\chi$ particles in general and hence we need to choose the conformal coupling $\xi=1/6$~\cite{Gorbunov:2010bn,Gorbunov:2012ns,Bernal:2020qyu,Li:2021fao}. 
However, the leading contribution from the inflaton coherent oscillation $\phi\phi\to\chi\chi$ is also cancelled out for the conformal coupling $\xi=1/6$.

\bibliographystyle{utphys}
\bibliography{ref}

\end{document}